\documentclass[proceedings]{JHEP3}

\PrHEP{PrHEP hep2001}                   
\conference{International Europhysics Conference on HEP}                

\title{Knot Theory from the Perspective of Field and String Theory}

\author{\speaker{Jose M. F. Labastida}\\               
        Departamento de F\'\i sica de Part\'\i
culas\\ Universidade de Santiago de Compostela\\
E-15706 Santiago de Compostela, SPAIN \\          
        E-mail: \email{labasti@fpaxp1.usc.es}}

\abstract{I present a summary of the recent progress made in field and string theory which
has led to a  reformulation of quantum-group polynomial invariants for knots and links into
new polynomial invariants whose coefficients can be described in topological terms. The
approach opens a new point of view in the theory of knot and link invariants.}

\begin{document}

\def\tr{{\hbox{\rm Tr}}}
\def\ex{{\hbox{\rm e}}}
\def\k{{\cal K}}

\newcommand{\beq}{\begin{equation}}
\newcommand{\eeq}{\end{equation}}
\newcommand{\bear}{\begin{eqnarray}}
\newcommand{\eear}{\end{eqnarray}}

\newcommand{\CS}{{\scriptstyle {\rm CS}}}
\newcommand{\CSs}{{\scriptscriptstyle {\rm CS}}}
\newcommand{\ie}{{\it i.e.}}

\newdimen\tableauside\tableauside=1.0ex
\newdimen\tableaurule\tableaurule=0.4pt
\newdimen\tableaustep
\def\phantomhrule#1{\hbox{\vbox to0pt{\hrule height\tableaurule width#1\vss}}}
\def\phantomvrule#1{\vbox{\hbox to0pt{\vrule width\tableaurule height#1\hss}}}
\def\sqr{\vbox{%
  \phantomhrule\tableaustep
  \hbox{\phantomvrule\tableaustep\kern\tableaustep\phantomvrule\tableaustep}%
  \hbox{\vbox{\phantomhrule\tableauside}\kern-\tableaurule}}}
\def\squares#1{\hbox{\count0=#1\noindent\loop\sqr
  \advance\count0 by-1 \ifnum\count0>0\repeat}}
\def\tableau#1{\vcenter{\offinterlineskip
  \tableaustep=\tableauside\advance\tableaustep by-\tableaurule
  \kern\normallineskip\hbox
    {\kern\normallineskip\vbox
      {\gettableau#1 0 }%
     \kern\normallineskip\kern\tableaurule}%
  \kern\normallineskip\kern\tableaurule}}
\def\gettableau#1 {\ifnum#1=0\let\next=\null\else
  \squares{#1}\let\next=\gettableau\fi\next}

\tableauside=1.0ex
\tableaurule=0.4pt


During the last years the theory of knot and link invariants has
experienced important progress. The confluence of Chern-Simons gauge theory and string
theory has led to a very powerful new approach which provides a topological interpretation
for the  integer coefficients of a reformulated version of quantum-group polynomial
invariants. The main goal of this short note is to present a summary of these recent
developments.


Chern-Simons gauge theory  is a
topological  quantum field theory whose action is built out of a
Chern-Simons term involving as gauge field a gauge connection associated
to a group $G$ on a three-manifold $M$. Its natural observables are Wilson loops, $
W^{K}_{R}$,  where $K$ is a loop and $R$ a representation of the gauge group. The
vacuum expectation values of products of these operators are topological
invariants which are related to quantum-group polynomial invariants. Given a link
${\cal L}$ of $L$ components, $K_1,K_2,\dots,K_L$, one computes
correlators of the form $\langle  W^{{K}_1}_{
R_1}\cdots W^{{K}_L}_{ R_L}\rangle $, where $R_1,R_2,\dots,R_L$ are
representations associated to each component. For $SU(N)$ as gauge group these quantities
turn out to be polynomials in $q=\ex^{2\pi i\over k+N}$ and $\lambda=q^N$ with
integer coefficients, being $k$ the Chern-Simons integer parameter. 

Witten found out in 1988
\cite{cs}, using non-perturbative methods,  that these correlators lead to
quantum-group polynomial invariants for knots and links \cite{qg}.
Perturbative methods have been also developed for
Chern-Simons gauge theory, lading to important connections related to Vassiliev invariants. A
summary of these developments can be found in a recent review
\cite{laplata}. In this short note I will concentrate in the new perspective
emerged after studying the large
$N$ expansion of Chern-Simons gauge theory. The discussion will be restricted to the case
of knots on $S^3$ with gauge group $SU(N)$.

Besides the perturbative
expansion, gauge theories with gauge group $SU(N)$ admit a large-$N$ expansion. In this
expansion correlators are expanded in powers of $1/N$ while keeping the 't Hooft coupling
$t=N x$ fixed, being
$x$  the coupling constant of the gauge theory. For example, for the
free energy of the theory one has the general form,
\begin{equation}
F=\sum_{g\ge 0 \atop h\ge 1}^\infty C_{g,h} N^{2-2g} t^{2g-2+h}.
\label{gw}
\end{equation}
In the case of Chern-Simons gauge theory, the coupling constant is
$x={2\pi i \over k+N}$ after taking into account the standard shift in $k$.
This large-$N$ expansion resembles a string
theory expansion and indeed the quantities $C_{g,h}$ can be identified
with the partition function of a topological open string with $g$ handles
and $h$ boundaries, with $N$ D-branes on $S^3$ in an ambient
six-dimensional target space $T^*S^3$. This was pointed out by Witten in
1992 \cite{witdb}. The result makes a connection between a topological
three-dimensional field theory and topological string theory.

An important breakthrough took place in 1998 after the discovery of a new
approach to compute the free energy (\ref{gw}). Using arguments
inspired by the AdS/CFT correspondence (see
\cite{ads} for a review), Gopakumar and Vafa
\cite{gova} provided a closed-string theory interpretation of the
free energy (\ref{gw}). They conjectured that it
can be expressed as
$
F=\sum_{g\ge 0 }^\infty  N^{2-2g} F_g(t),
$
where $F_g(t)$ corresponds to the partition function of a topological
closed string theory on the non-compact Calabi-Yau manifold $X$ called
the resolved conifold,
${\cal O}(-1) \oplus {\cal O}(-1) \rightarrow {\bf P}^1,$
being $t$ the flux of the $B$-field through ${\bf P}^1$.
The quantities $F_g(t)$ have been computed using physical \cite{gova} and 
mathematical arguments \cite{rig}, proving the conjecture.

The study of observables in this new context
was first faced by Ooguri and Vafa, and they provided the appropriate picture \cite{oova}
which was later refined in \cite{lmv}. 
To consider the presence of Wilson loops it is convenient to introduce a
particular generating functional. First, one performs a change
of basis from representations $R$ to conjugacy classes $C(\vec k)$ of the
symmetric group, labeled by vectors $\vec k= (k_1,k_2,\dots)$ with $k_i\ge
0$, and
$|\vec k| = \sum_j k_j > 0$. The change of basis is $W_{\vec k} = \sum_R
\chi_R(C(\vec k)) W_R$, where $\chi_R$ are characters of the permutation
group $S_\ell$ of
$\ell=\sum_j j k_j$ elements ($\ell$ is also the number of boxes of the
Young tableau associated to $R$). Second, one introduces
the generating functional, $F(  V )=\log Z(  V  )= \sum_{\vec k}
{|C({\vec k} )|
\over   \ell  !}  W^{  (c)}_{{\vec k} } 
\Upsilon_{{\vec k}}(  V )$,
where $Z(  V  )= \sum_{\vec k}
{|C({\vec k} )|
\over   \ell  !}  W^{}_{{\vec k} } 
\Upsilon_{{\vec k}}(  V )$ and $\Upsilon_{\vec{k}}(
V)=\prod_j ({\rm Tr} V^{ j})^{k_{ j}} $. In these expressions $|C({\vec
k} )|$ denotes the number of elements of the class $C({\vec k} )$ in
$S_\ell$. The reason behind the introduction of this generating
functional is that the large-$N$ structure of the connected Wilson
loops, $W^{  (c)}_{\vec{k} }$, turns out to be very simple:
\begin{equation}
{  |C(\vec{k} )| \over  \ell  !}
W^{  (c)}_{\vec{k} } =\sum_{g=0}^{\infty}  x^{  2g-2+
|\vec{k} |} 
F_{g, \vec{k} }(\lambda ),
\label{largenwil}
\end{equation}
where $\lambda = \ex^t$ and $t=N x$ is the 't Hooft coupling. Writing
$x=t/N$, it corresponds to a power series expansion in $1/N$. As before,
the expansion looks like a perturbative series in string theory where $g$
is the genus and $|\vec k|$ is the number of holes. Ooguri and Vafa
conjectured in 1999 the appropriate string theory description of
(\ref{largenwil}). It corresponds to an open topological string theory
whose
target space is the resolved conifold $X$. The contribution from this
theory leads to open-string analogs of Gromov-Witten invariants.

To describe in detail the fact that one is dealing with open
strings, some new data needs to be introduced. Here is where the knot
intrinsic to the Wilson loop enters. Given a knot $K$ on $S^3$, let us
associate to it a Lagrangian submanifold $C_K$ with $b_1=1$ in the resolved
conifold $X$ and consider a topological open string on it. The
contributions in this open topological string are localized on holomorphic
maps $f:\Sigma_{g,h}
\rightarrow X$ with $h=|\vec k|$ which satisfy: $f_*[\Sigma_{g,h}]={\cal
Q}$, and $f_*[C]=j[\gamma]$ for $k_j$ oriented circles $C$. In these
expressions $\gamma\in H_1(C_K,{\bf Z})$, and ${\cal Q}\in H_2(X,C_K,{\bf
Z})$, \ie, the map is such that $k_j$ boundaries of $\Sigma_{g,h}$ wrap
the knot $j$ times, and $\Sigma_{g,h}$ itself gets mapped to a relative
two-homology class characterized by the Lagrangian submanifold $C_K$.
The number of these maps (defined in an appropriate form)
constitutes the open-string analog of Gromov-Witten invariants. They will
be denoted by $N_{g,\vec k}^{\cal Q}$. The quantity $F_{g, \vec{k} }(\lambda )$ in
(\ref{largenwil}) takes the form:
\begin{equation}
F_{g, \vec{k} }(\lambda ) = \sum_{\cal Q} N_{g,\vec k}^{\cal Q}
\ex^{\int_{\cal Q} \omega}, \,\,\,\,\,\,\,\,\,\,\,\,\,\,\,\,
t=\int_{{\bf P}^1} \omega,
\label{gwopen}
\end{equation}
where $\omega$ is the  K\"ahler class of the Calabi-Yau manifold $X$ and
$\lambda = \ex^t$. For any ${\cal Q}$, one can always write $\int_{\cal
Q} \omega = Qt$ where $Q$ is in general a half-integer number. Therefore,
$F_{g, \vec{k} }(\lambda )$ is a polynomial in $\lambda^{\pm {1\over 2}}$
with rational coefficients.

The result (\ref{gwopen}) constitutes a first step to provide a
representation where one can assign a geometrical interpretation to the
integer coefficients of the quantum-group invariants.
Notice that to match a polynomial invariant to 
(\ref{gwopen}), after
obtaining its connected part, one must expand  it in $x$ after setting
$q=\ex^x$ keeping $\lambda$ fixed. One would like to have a refined version
of (\ref{gwopen}).
This study was indeed done in \cite{oova} and later
improved in \cite{lmv}. The outcome is that $F(V)$ can be
expressed in terms of integer invariants related to topological strings.

To present these results one needs to perform first a
reformulation of the quantum-group invariants or vacuum expectation
values of Wilson loops. Instead of considering $W_R(q,\lambda)$, a
corrected version of it, $f_R(q,\lambda)$, will be studied. These
reformulated polynomial invariants have the form:
\begin{eqnarray}
 f_{ R }(q, \lambda) &=&
 \sum_{d, m=1}^{\infty} (-1)^{m-1} {\mu 
(d) \over d m} 
\sum_{ \{ \vec{k}^{(j)},  R_{ j} \} } 
 \chi_{ R} \biggl( 
C\biggl( (\sum_{j=1}^m \vec{k}^{(j)})_d\biggr)\biggr)\nonumber\\
&& \times \prod_{j=1}^m {|C(\vec{k}^{(j)})| \over  \ell_{
 j}!}
  \chi_{ R_{ j}}(C(\vec{k}^{(j)})) 
W_{ R_{ j}}(q^{ d}, \lambda^{ d}),
\label{lafor}
\end{eqnarray}
where $({\vec k}_d)_{di} = k_i$ and zero otherwise. In this expression
$\mu(d)$ is the Moebius function. The
reformulated polynomial invariants $f_R(q,\lambda)$ are just
quantum-group invariants $W_R(q,\lambda)$ plus lower order terms,
understanding by this terms which contain quantum-group invariants
carrying representations whose associated Young tableaux have a lower
number of boxes. For example, for the simplest cases:
\begin{eqnarray}
f_{\tableau{1}}(q,\lambda)&=&W_{\tableau{1}}(q, \lambda), \nonumber\\
f_{\tableau{2}}(q,\lambda)&=&W_{\tableau{2}}(q,\lambda)
-{1\over 2}\bigl( W_{\tableau{1}}(q,\lambda)^2+ 
W_{\tableau{1}}(q^2,\lambda^2)
\bigr),\nonumber\\
f_{\tableau{1 1}}(q, \lambda)&=&W_{\tableau{1 1}}(q,\lambda)
-{1\over 2}\bigl( W_{\tableau{1}}(q,\lambda)^2-
 W_{\tableau{1}}(q^2,\lambda^2) 
\bigr).
\label{examples}
\end{eqnarray}
The nice feature of the reformulated quantities is that $F(V)$
acquires a very simple form in terms of them:
\begin{equation}
F(V)=\sum_{d=1}^\infty \sum_R {1\over d} f_R(q^d,\lambda^d) {\rm Tr} V^d,
\label{simple}
\end{equation}
and therefore they seem to be the right quantities to express an
alternative point of view in which the embedded Riemann surfaces can be
regarded as D-branes.

Two more ingredients are needed to present the conjectured form of the reformulated
$f_R(q,\lambda)$. One needs the Clebsch-Gordon
coefficients $C_{R,R',R''}$ of the symmetric group (they satisfy
$V_R\otimes V_{R'} = \sum_{R''} C_{R,R',R''} V_{R''}$), and  monomials
$S_R(q)$ defined as follows: $S_R(q)=(-1)^d q^{d-{\ell-1\over 2}}$ if $R$
is a hook representation, with $\ell-d$ boxes in the first row, and
$S_R(q)=0$ otherwise. The conjecture presented in
\cite{lmv} states that the reformulated invariants have the form:
\begin{equation}
f_{  R}  (q ,
\lambda )=\sum_{g\ge 0}\sum_{Q,  R' ,   R''} 
C_{R R' R''}  N_{  R' , g, Q} S_{ 
R'' }(q )(q^{ {1
\over 2}} - q ^{  -{1
\over 2}})^{2g-1} \lambda^{  Q},
\label{conjecture}
\end{equation}
where $N_{  R , g, Q}$ are {\it integer} invariants which posses a geometric
interpretation. These quantities are alternative integers
 in Gromov-Witten theory. They can be described in terms of
the moduli space of Riemann surfaces with boundaries embedded into a
Calabi-Yau manifold. Their geometrical interpretation has been treated
recently in \cite{last}. 

The structure present in (\ref{conjecture}) has been verified for a variety of
non-trivial knots and links, and representations up to four boxes
\cite{testing,lmv}. For the unknot, the whole picture have been verified in
complete detail \cite{oova,nonudo}. The form of $F(V)$ for the unknot can be
easily computed in Chern-Simons gauge theory,
\begin{equation}
  F(  V )=\sum_{d=1}^{\infty} { \lambda^{d  \over2} - 
\lambda^{-{d \over 2}} 
\over 2d \sin \bigl( {d x   \over 2} \bigr)} {\rm
Tr}_{ \tableau{1}} {  V }^d,
\label{unknot}
\end{equation}
leading to an expansion (\ref{largenwil}) of the form:
\begin{equation}
F_{g, (  0  ,\cdots,   0 ,  1 ,
  0 ,
\cdots,
  0 )}(\lambda)={ (1-2^{1-2g})|B_{2g}|
\over  (2g)! }d^{2g-2} (\lambda^{d  \over2} - 
\lambda^{-{d \over 2}}), 
\label{bern}
\end{equation}
where  the 1 in $F_{g, (  0  ,\cdots,   0 ,  1 ,
  0 ,
\cdots,
  0 )}$ is located in the $d^{\rm th}$ position. Form these equations one can
easily read the numbers which correspond to the open-string analogs of
Gromov-Witten invariants,
$N_{g,\vec k}^{\cal Q}$, in (\ref{gwopen}), as well as the new integer
invariants present in the general expression (\ref{conjecture}):
$
N_{\tableau{1},0,{1\over 2}} = - N_{\tableau{1},0,-{1\over 2}}=1.
$

Many issues remain open in this context and much work is being carried out to
unravel the consequences of this new connection between theoretical physics and topology. I
hope that in this short note I have convinced the reader of the emergence of a fascinating
interplay between string theory, knot theory and enumerative geometry which opens new fields
of study.


I would like to thank the organizers of the International Europhysics Conference on High
Energy Physics for inviting me to deliver a talk. I would like to
thank also  M. Mari\~no  for collaborations and discussions on many of the
topics described here. This work is supported in part by Ministerio de Ciencia y
Tecnolog\'\i a under grant PB96-0960, and by Xunta de Galicia under grant
PGIDT00-PXI-20609.



\end{document}